# Gravitational effectiveness of the zero-point energy of the radiation field: a possible solution of a paradox raised by Pauli

P. R. Silva

Departamento de Física – ICEx – Universidade Federal de Minas Gerais
C. P. 702 - 30123-970 – Belo Horizonte – MG - Brasil
e-mail:prsilvafis@terra.com.br

ABSTRACT- A modified vacuum energy density of the radiation field is evaluated, which leads to accepted prediction for the radius of the universe. The modification takes into account the existence of a new gauge boson which also can be used in order to determine the mass of the boson responsible for the weak decay of the muon.

## 1 - INTRODUCTION

The cosmological constant problem has been a subject of great interest in the last years, as can be verified in the work of Nobbenhuis [1]. (Please see also references cited therein). In a historical account of the cosmological constant problem (the $\Lambda$-problem) Straumann [2] states that in the early years of the quantum mechanics (1920) Pauli wondered whether the zero-point energy of the radiation field could be gravitationally effective. As pointed out by Straumann: from Pauli's discussion with Enz and Thellung [2] we know that Pauli estimated the influence of the zero-point energy of the radiation field – cut off at the classical electron radius - on the radius of the universe and came to the conclusion that it "could not even reach to the moon".

In this work we intend to modify the simple calculations presented by Straumann as a means to obtain the radius of the universe as a function of the fine structure constant, the electron mass, and the Planck mass. As we will see, the radius of the universe estimated in this work agrees in order of magnitude with its current accepted value [8]. The key ingredient of the present calculations will be a proposed existence of a new gauge boson which intermediates the transitions, in this way modifying the energies of the radiation field. This boson will be interpreted with that was introduced by Zee [3] as responsible by a "new physics".

## 2 - THE NEW GAUGE BOSON

In a recent paper [4] the time evolution of the universe world line was compared with the growing of a polymer chain. The lattice spacing was taking as the Planck length $L_P$ and the radius of gyration $R_\Lambda$ was related to the Compton wavelength of a particle of mass $M_\Lambda$ associated to the cosmological constant or dark energy.

It would be interesting to look for a polymer which chain has an extension equals to the Compton wavelength of a particle of mass equals to that of an electron-positron pair. We will consider the lattice spacing of this chain having the Planck length which naturally sets up the ultraviolet cut off of the problem.

Let us write the Flory's free energy [4,5] of the pair as

$$F_{pair}/t = (N^2 L_P^d)/R^d + R^2/(N L_P^2), \qquad (1)$$

where t is an unspecified temperature, and N the number of monomers in the chain.

Equation (1) is the equivalent of equation (9) of reference [4]. Setting $L = NL_P$ and minimizing (1) relative to R, we obtain the radius of gyration of the chain related to the electron-positron pair, namely we get

$$R_{gyr} = L^{3/(2+d)} L_P^{(d-1)/(2+d)}. \tag{2}$$

In the above relation, d is the space-time dimension. We are particularly interested in the d=4 dimension. We have

$$R_{gyr.}|_{d=4} = L^{1/2} L_P^{1/2}. \tag{3}$$

If we define

$$M_{pair} = 2\,m_e = L^{-1} \quad \text{and} \quad M_B = (R_{gyr}|_{d=4})^{-1}, \tag{4}$$

where $m_e$ is the electron mass and $M_B$ is the mass of the gauge boson, we get

$$M_B = (2m_e\,M_P)^{1/2}. \tag{5}$$

We will interpret $M_B$ given by equation (5) as the mass of a boson responsible by the "new physics" as conjectured by Hsu and Zee [3].
Taking $2m_e = 10^{-3}$ GeV and $M_P = 10^{19}$ GeV, we obtain

$$M_B = 10^8 \text{ GeV}. \tag{6}$$

It is also interesting to evaluate N, the number of monomers in the chain. We have

$$N = L/L_P = M_P/2m_e = 10^{22}. \tag{7}$$

We also have

$$R_{gyr}|_{d=4} = N^{1/2} L_P = 10^{-24} \text{ m}. \tag{8}$$

3 - ESTIMATION OF THE RADIUS OF THE UNIVERSE

Now let us consider a process related to the creation or annihilation of an electron-positron pair which is intermediated by a boson of mass $M_B$. Next we will present a non-linear wave equation describing the scattering of an electron-positron pair where the interaction strength is given by the fine structure constant alpha and its range goes like the inverse of the boson mass. The situation is similar to that found in the weak decay process. We write

$$\Delta \Psi - \partial^2 \Psi / \partial t^2 = 4m_e^2 \Psi - \alpha M_B^2 \Psi^2. \tag{9}$$

We have written an equation similar to (9) in a paper dealing with the proton lifetime. Please see reference [6]. In (9) we took $\hbar = c = 1$.
A "stationary" solution of (9) gives

$$0 = 4m_e^2 \Psi - \alpha M_B^2 \Psi^2. \tag{10}$$

Solving (10) for Ψ leads to

$$\Psi = (4m_e^2)/(\alpha M_B^2). \tag{11}$$

Taking into account all the possible transitions of energy ω weighted by $\Psi^2$, we have

$$\Gamma = \omega \Psi^2 = (\omega/\alpha^2)(2m_e/M_B)^4 = \omega_{mod}, \tag{12}$$

where $\omega_{mod}$ means "modified ω".

Usually the vacuum energy density of the radiation field is given by

$$<\rho>_{vac} = 8/(2\pi)^3 \int_o^{\omega max} (\omega/2) \omega^2 d\omega = (1/8\pi^2) \omega_{max}^4. \tag{13}$$

In (13) $\omega_{max}$ is fixed by considering the cut off provided by the classical radius of the electron, namely

$$\omega_{max} = (2\pi)/\lambda_{max} = 1/R_{Cl,e} = m_e/\alpha. \tag{14}$$

On the other hand the modified vacuum energy density reads

$$<\rho>_{mod} = 8/(2\pi)^3 \int_o^{\omega max} \tfrac{1}{2}(\omega_{mod}) \omega^2 d\omega = (1/8\pi^2)(1/\alpha^2)(2m_e/M_B)^4 \omega_{max}^4. \tag{15}$$

Inserting (14) into (15) we get

$$<\rho>_{mod} = (1/8\pi^2)(2m_e/M_B)^4(m_e^4/\alpha^6). \tag{16}$$

In the presence of the cosmological constant term the field equations in the standard notation (please see Straumann [7]) and signature (+---) are

$$G_{\mu\upsilon} = 8\pi G\, T_{\mu\upsilon} + \Lambda\, g_{\mu\upsilon}. \tag{17}$$

For the static Einstein universe the field equations (17) imply the two relations

$$8\pi G \rho = 1/a^2 = \Lambda, \tag{18}$$

where ρ is the energy density and a the radius of curvature.

Inserting $<\rho>_{mod}$, given by (16), in the place of ρ of (18) and solving for a, we have

$$a = (8\pi G <\rho>_{mod})^{-1/2} = (2\alpha^3/m_e)[M_P/(2m_e)]^2. \tag{19}$$

In obtaining (19) we have used (16) and (5) and also that $M_P = G^{-1/2}$.
Evaluating the right side of (19) we obtain a radius of the universe of order of magnitude $10^{25}$ m which is close to its current accepted value [8].

4 - THE "NEW PHYSICS" AND THE MUON MASS

In reference [3], a boson responsible for a "new physics" was associated to a muon decay process. Thinking in terms of the weak interaction it is possible to write for the muon mass the relation

$$m_\mu = \alpha_{w,\mu}/(2R_\mu), \tag{20}$$

where the weak coupling "constant", $\alpha_{w,\mu}$, is evaluated at the muon mass-energy, but the range of the interaction $R_\mu$ is related to the mass of the new boson. By considering

$$\alpha_{w,\mu} = \alpha(m_\mu/M_w)^2, \quad \text{and} \quad \tfrac{1}{2} M_B = (R_\mu)^{-1}, \tag{21}$$

we have

$$m_\mu = (\alpha/4)(m_\mu/M_w)^2 M_B. \tag{22}$$

Using (5) and solving (22) for $M_w$, we finally obtain

$$M_w = \tfrac{1}{2}(\alpha\, m_\mu)^{1/2}(2m_e M_P)^{1/4}. \tag{23}$$

Therefore we were able to write the mass of the gauge boson responsible for the weak decay in terms of electron, muon and Planck masses, besides the electromagnetic coupling. Putting these numbers in (22) we find for $M_w$ an order of magnitude estimation of $10^2$ GeV.

5 - DISCUSSION

In this work we evaluated the radius of the universe and the mass of the gauge boson of the weak interaction as functions of the masses of some elementary particles, the electromagnetic coupling and the Planck mass. A new gauge boson played a fundamental role in the determination of these quantities. In order to evaluate the mass of this boson we considered polymers models, where the monomers which compose the chains have size equals to the Planck length. Therefore consideration of the discreteness of the space-time seems to be very relevant to accomplish this task.

ACKNOWLEDGEMENT

We are grateful to colleague Nilton Penha Silva for reading the manuscript.

REFERENCES

[1] Stefan Nobbenhuis, Ph.D. Thesis, arXiv:gr-qc/0609011v1  4 Sep 2006.
[2] N. Straumann, arXiv:gr-qc/0208027v1   13Aug 2002.
[3] A. Zee, in:Quantum Field Theory in a Nutshell, Princeton University Press, 2003, page 441.
[4] P. R. Silva, Braz. J. Phys. **38,** 587(2008).
[5] P-G de Gennes, Scaling Concepts in Polymer Physics, Chap. I, Cornell University Press, Ithaca, N. Y.,(1979).
[6] P. R. Silva, Mod Phys. Letters A, **10**, 911(1995).
[7] N. Straumann, General Relativity and Relativistic Astrophysics, Spring-Verlag(1984).
[8] P. R. Silva, Int. Jour. Mod. Phys.**A, 12,**1373(1997)**.**